\documentclass[prl,aps,twocolumn,showpacs,amsmath,amssymb]{revtex4}
\usepackage{graphicx}% Include figure files
\usepackage{dcolumn}% Align table columns on decimal point
\usepackage{bm}% bold math

\begin{document}
\title{Giant isotope effect and spin state transition induced by oxygen isotope
exchange in ($Pr_{1-x}Sm_x)_{0.7}Ca_{0.3}CoO_3$}

\author{ G. Y. Wang}
\author{X. H. Chen}
\altaffiliation{Corresponding author} \email{chenxh@ustc.edu.cn}
\author{ T. Wu, G. Wu, X. G. Luo and C. H. Wang}
\affiliation{Hefei National Laboratory for Physical Science at
Microscale and Department of Physics, University of Science and
Technology of China, Hefei, Anhui 230026, People's Republic of
China\\ }

\date{\today}

\begin{abstract}
We systematically investigate effect of oxygen isotope in
$(Pr_{1-x}Sm_x)_{0.7}Ca_{0.3}CoO_3$ which shows a crossover with x
from ferromagnetic metal to the insulator with spin-state
transition. A striking feature is that effect of oxygen isotope on
the ferromagnetic transition is negligible in the metallic phase,
while replacing $^{16}O$ with $^{18}O$ leads to a giant up-shift
of the spin-state transition temperature ($T_s$) in the insulating
phase, especially $T_s$ shifts from 36 to 54 K with isotope
component $\alpha_S=-4.7$ for the sample with x=0.175. A
metal-insulator transition is induced by oxygen isotope exchange
in the sample x=0.172 being close to the insulating phase. The
contrasting behaviors observed in the two phases can be well
explained by occurrence of static Jahn-Teller distortions in the
insulating phase, while absence of them in the metallic phase.
\end{abstract}


\maketitle
\newpage
Cobaltites with $CoO_6$ octahedra have recently attracted much
attention because they exhibit various intriguing physical
properties as magnetoresistance,\cite{Briceno} enormous Hall
effect,\cite{Samoilov} superconductivity,\cite{takada} and large
thermoelectric effect.\cite{Terasaki}  Provskite-related cobalt
oxides show their unique feature: temperature-induced spin state
transition (SST).\cite{Senaris,Tsubouchi} Such spin state
transition, which is extremely sensitive to composition and
pressure,\cite{Asai} is closely connected with electrical
conductivity. The $Co^{3+}$ ion, which contains six 3d electrons,
often exhibits a SST from the low spin (LS, $t_{2g}^6$) ground
state to the intermediate spin (IS, $t_{2g}^5 e_g^1$) or to the
high spin (HS, $t_{2g}^4e_g^2$) state with increasing temperature.
The most important character of the SST in perovskite-type cobalt
oxides is often accompanied by the metal-insulator transition
(MIT).

Experimental and theoretical works indicated that the spin-state
transition in cobaltites is from LS to IS below room
temperature.\cite{Tsubouchi,louca,saitoh,korotin} The IS state is
expected to be strong Jahn-Teller (JT) active due to the partially
filled $e_g$ level, creating monoclinic distortions of $CoO_6$
octahedra. Actually, elastic and inelastic neutron scattering
showed that spin activation of $Co^{+3}$ ions in
$La_{1-x}Sr_xCoO_3$ system induces local static JT distortions in
the paramagnetic insulating phase for all x, while the static JT
distortions are absent in the ferromagnetic metal
phase.\cite{louca1} Understanding the characteristics of
Jahn-Teller instability can help elucidate the nature of
metal-insulator transition, magnetoresistance, and even
superconductivity.\cite{bersuker}  The IS-JT state is stabilized
with doping in the insulating phase for perovskite cobalt
oxides.\cite{louca2,Tsubouchi} However, charge dynamics interfere
with orbital JT ordering resulting in a new state and the JT
distortions can lose their long-range coherency with increasing
charge mobility in the ferromagnetic metal phase. JT interactions
in perovskite oxides often result in quite intriguing physical
phenomena through the coupling between orbital and spin degrees of
freedom with lattice.\cite{goodenough} Coupling of JT modes to the
lattice results in formation of small JT polarons, so that
coupling of the charge carrier to the JT lattice distortions leads
to a giant oxygen isotope effect, and the JT distortions play an
essential role in $La_{1-x}Ca_xMnO_{3+y}$.\cite{zhao} Especially,
a metal-insulator transition was induced by oxygen isotope
exchange in $La_{0.175}Pr_{0.525}Ca_{0.3}MnO_3$.\cite{Babushkina}
To investigate JT physics and its role played in cobaltites, here
we systematically study effect of oxygen isotope in perovskite
$(Pr_{1-x}Sm_x)_{0.7}Ca_{0.3}CoO_3$ system with different x. This
system is suitable to study the JT physics by oxygen isotope
effect since it shows a crossover at x$\sim$0.175 from a
\emph{ferromagnetic metal phase} to \emph{an insulating phase}
with SST with increasing x.\cite{fujita, fujita1} Therefore, it is
expected that effect of oxygen isotope should be quite different
in the ferromagnetic metal phase from in the insulating phase
because the static JT distortions are observed in the insulating
phase, while absent in the metallic phase.

Polycrystalline samples $(Pr_{1-x}Sm_x)_{0.7}Ca_{0.3}CoO_3$ with
different x were prepared by conventional solid-state reaction.
Proper molar ratios of $Pr_6O_{11}$, $Sm_2O_3$, $CaCO_3$ and
$Co_3O_4$ were mixed, well ground and then calcined at 1000 $^o$C
and 1100 $^oC$ for 24 h with intermittent grinding. The pellets
pressed from the powder were sintered at 1200 $^oC$ in the flowing
oxygen for 48 h. All samples were characterized by X-ray
diffraction and found to be single phase. One pellet for the
samples with different x obtained above was cut into two pieces
for oxygen-isotope diffusion. The two pieces for each composition
were put into an alumina boat which were sealed in a quartz tube
filled with oxygen pressure of 1 bar (one for $^{16}O$ and another
for $^{18}O$) mounted in a furnace, respectively. The quartz tubes
formed parts of two identical closed loops. They were heated at
1000 $^oC$ for 48 h and then slowly cooled to room temperature.
The obtained samples were re-examined by X-ray diffraction to
confirm them single phase. The oxygen-isotope enrichment is
determined from the weight changes of both $^{16}O$ and $^{18}O$
samples. The $^{18}O$ samples have about 70\% $^{18}O$ and 30\%
$^{16}O$. To make sure the isotope exchange effect, back-exchange
of $^{18}O$ sample by $^{16}O$ was carried out in the same way and
the weight change showed a complete back-exchange. Magnetic
susceptibility measurements were performed with a SQUID
magnetometer in a magnetic field of 1000 Oe. Resistance
measurements were performed using the ac four-probe method with an
ac resistance bridge system (Linear Research Inc. LR-700P).

\begin{figure}[b]
\includegraphics[width=9cm]{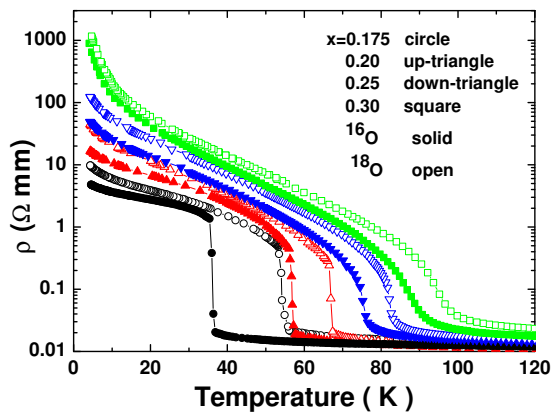}
\caption{\label{fig:epsart}  Temperature dependence of resistivity
for the $^{16}O$ and $^{18}O$ samples
$(Pr_{1-x}Sm_x)_{0.7}Ca_{0.3}CoO_3$ with x=0.175, 0.20, 0.25
 and 0.30. The data denoted by the solid
and open symbols are taken from the samples treated in $^{16}O$
and $^{18}O$, respectively. }
\end{figure}

Figure 1 shows temperature dependence of resistivity for the
samples $(Pr_{1-x}Sm_x)_{0.7}Ca_{0.3}CoO_3$ with x=0.175, 0.20,
0.25 and 0.30 treated in $^{16}O$ and $^{18}O$, respectively. The
samples $(Pr_{1-x}Sm_x)_{0.7}Ca_{0.3}CoO_3$ with various x treated
in $^{16}O$ shows a very weak semiconductive behavior at the high
temperatures, while a sharp metal-insulator transition is observed
at certain temperatures ($T_S$) for different x with decreasing
temperature. $T_S$ increases with increasing x. As shown in Fig.1,
substitution of $^{18}O$ for $^{16}O$ leads to an enhancement of
$T_S$ for all samples. The up-shift of $T_S$ decreases with
increasing x. A giant isotope shift is observed in the sample with
x=0.175. $T_S$ shifts from 36 to 54 K after replacing of $^{16}O$
by $^{18}O$. The oxygen isotope exponent $\alpha_s = -
dlnT_S/dlnM_O$ (where $M_O$ is the oxygen isotope mass) is about
-4.7, which is much larger than $\alpha_ c=0.8$ observed in
$La_{1-x}Ca_xMnO_{3+y}$.\cite{zhao} However, the up-shift of $T_S$
is about 7 K for the sample with x=0.30 and the isotope component
$\alpha_s$ is about -0.8. To make sure that the effect arises from
the oxygen isotope exchange, we carried out the isotope
back-exchange. The transport properties completely return to that
observed in the samples previously treated in $^{16}O$ and
$^{18}O$ after the isotope back-exchange, respectively. Such
metal-insulator transition has been believed to arise from the
spin-state transition. Therefore, the susceptibility was measured
for the samples above.

\begin{figure}[b]
\includegraphics[width=9cm]{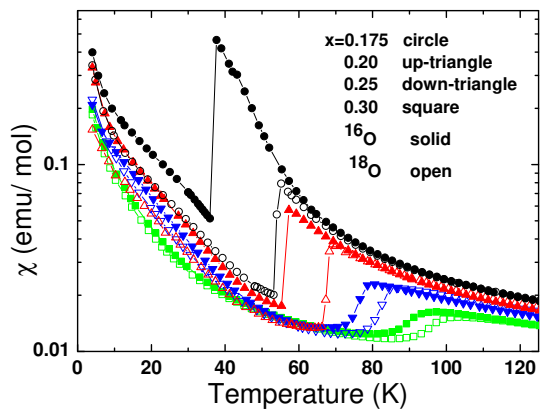}
\caption{\label{fig:epsart}   Temperature dependence of
susceptibility for the $^{16}O$ and $^{18}O$ samples
$(Pr_{1-x}Sm_x)_{0.7}Ca_{0.3}CoO_3$ with x=0.175, 0.20, 0.25
 and 0.30. The data denoted by the solid and open symbols are
taken from the samples treated in $^{16}O$ and $^{18}O$,
respectively.}
\end{figure}

Temperature dependence of susceptibility measured in the
field-cooled process with magnetic field of 1000 Oe is shown in
Fig.2 for the same samples used in resistivity measurements. All
samples show a sharp SST from IS state to LS state with decreasing
temperature, respectively. $T_s$ increases with increasing x.
Replacing of $^{16}O$ with $^{18}O$ makes $T_s$ shift to
high-temperature. This is consistent with metal-insulator
transition observed in resistivity as shown in Fig.1. It should be
pointed out that $T_s$ is exactly the same as that of
metal-insulator transition observed in Fig.1. The sample with
x=0.175 treated in $^{16}O$ and $^{18}O$ show a sharp SST from IS
state to LS state with decreasing temperature at 36 and 54 K,
respectively. It further supports that the metal-insulator
transition arises from the SST. It has been noted that there
exists a thermal hysteresis for the phase transition. Therefore,
all measurements were carried out in the cooled process to keep
the consistency.

\begin{figure}[h]
\includegraphics[width=9cm]{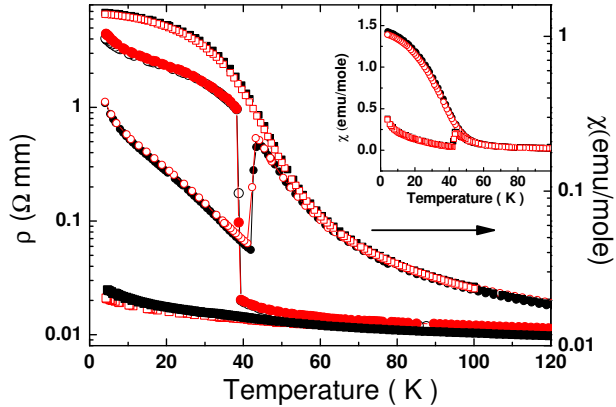}
\caption{\label{fig:epsart}Temperature dependences of resistivity
and susceptibility for the sample
$(Pr_{1-x}Sm_x)_{0.7}Ca_{0.3}CoO_3$ with x=0.172 (solid squares
for the $^{16}O$ sample; solid circles for $^{18}O$ sample). The
data denoted by the open symbols were obtained after the oxygen
isotope back-exchange ($^{16}O\rightarrow^{18}O$ and $^{18}O
\rightarrow^{16}O$). The reversibility of the metal-insulator
transition induced by oxygen isotope exchange is clear.}
\end{figure}

Figure 3 shows temperature dependences of resistivity and
susceptibility for the $^{16}O$ and $^{18}O$ samples
$(Pr_{1-x}Sm_x)_{0.7}Ca_{0.3}CoO_3$ with x=0.172. The sample
treated in $^{16}O$ shows a very weak semiconductive behavior at
the low temperatures and no phase transition, while a sharp
metal-insulator transition is observed at about 39 K after the
treatment in $^{18}O$. It suggests that a metal-insulator
transition is induced by oxygen isotope exchange.
 To make sure that the effect arises from
the oxygen isotope exchange, we carried out the isotope
back-exchange. As shown in Fig.3, the transport properties
completely return to that observed in the samples previously
treated in $^{16}O$ and $^{18}O$ after the isotope back-exchange,
respectively. The remarkable feature is that the metal-insulator
transition disappears after the subsequent treatment in $^{16}O$,
while shows up after the subsequent treatment in $^{18}O$. It
definitely indicates that the metal-insulator transition is
induced by replacing $^{16}O$ with $^{18}O$.

As shown in Fig.3, a striking feature for susceptibility is that
the sample $(Pr_{1-x}Sm_x)_{0.7}Ca_{0.3}CoO_3$ with x=0.172
treated in $^{16}O$ shows a ferromagnetic transition at about 60
K, while the ferromagnetic transition disappears and a sharp
spin-state transition at about 42 K is induced after replacing of
$^{16}O$ with $^{18}O$. To clearly show a ferromagnetic
transition, the same data are plotted with linear vertical axis in
the inset of Fig.3. It indicates that the replacement of $^{16}O$
with $^{18}O$ suppresses the ferromagnetic phase, while induces a
SST in the sample with x=0.172. It suggests that there exists a
competition between the ferromagnetic transition and SST. This
result is consistent with that observed in resistivity
measurements. It should be pointed out that the transition
temperature in susceptibility is about 3 K higher than that in
resistivity, in contrast to the case for samples with $x\geq
0.175$. The isotope back-exchange results further confirm that the
behavior observed in Fig.3 is due to oxygen isotope exchange.
Beyond the expectation, a phase transition from ferromagnetic
metal to an insulator with spin-state transition is induced by
replacing replacing $^{16}O$ with $^{18}O$.

\begin{figure}[h]
\includegraphics[width=9cm]{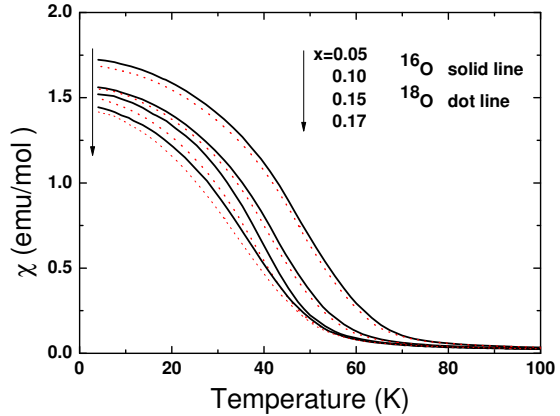}
\caption{\label{fig:epsart} Temperature dependence of
susceptibility for the $^{16}O$ and $^{18}O$ samples
$(Pr_{1-x}Sm_x)_{0.7}Ca_{0.3}CoO_3$ with x=0.05, 0.10, 0.15 and
0.17: solid lines for $^{16}O$ samples; dot lines for $^{18}O$
samples. }
\end{figure}

To further study the JT physics by oxygen isotope effect in the
ferromagnetic metal phase, the susceptibility is systematically
studied on the $^{16}O$ and $^{18}O$ samples
$(Pr_{1-x}Sm_x)_{0.7}Ca_{0.3}CoO_3$ with x=0.05, 0.10, 0.15 and
0.17. Temperature dependence of susceptibility for these samples
are shown in Fig.4. Actually, substitution of $^{18}O$ for
$^{16}O$ suppresses the ferromagnetic transition and leads to a
down-shift of the Curie temperature ($T_c$), being similar to that
observed in mamganites although the down-shift of $T_c$ is much
less relative to $La_{1-x}Ca_xMnO_{3+y}$.\cite{zhao} The
down-shift of $T_c$ slightly increases with increasing x. However,
the shift of $T_c$ is very small compared to that of $T_S$. It
suggests that the effect of oxygen isotope on ferromagnetic
transition is negligible.

In order to compare the effect of oxygen isotope in the
ferromagnetic metallic phase and in the insulating phase with
spin-state transition, the isotope components $\alpha_s$ for SST
and $\alpha_c$ for ferromagnetic transition as a function of x and
the average ionic radius of A site are shown in Fig.5. The isotope
component $\alpha_s$ decreases with increasing Sm content x in
$(Pr_{1-x}Sm_x)_{0.7}Ca_{0.3}CoO_3$ system. As observed in Fig.1
and Fig.2, $T_S$ increases with increasing x, while its isotope
shift $\Delta T_s$ decreases. Therefore, it leads to a decrease of
the isotope component $\alpha_s$ with increasing x. Substitution
of $^{16}O$ by $^{18}O$ leads to a decrease in the Curie
temperature ($T_c$) although its oxygen isotope shift is
negligible. Its isotope component $\alpha_c$ is positive and
increases with increasing x, in contrast to the $\alpha_s$. It
further confirms that there exists a competition between
ferromagnetic transition and SST. As shown in Fig.5, a maximum
isotope component $\alpha$ is observed in the boundary between
ferromagnetic metal and insulating phases. A intriguing phenomenon
is that the substitution of $^{16}O$ by $^{18}O$ leads to a
disappearance of ferromagnetic transition and induces the SST
accompanied by the metal-insulator transition in the sample
$(Pr_{0.828}Sm_{0.172})_{0.7}Ca_{0.3}CoO_3$ which is in the
ferromagnetic metal phase close to the crossover boundary.

\begin{figure}[t]
\includegraphics[width=9cm]{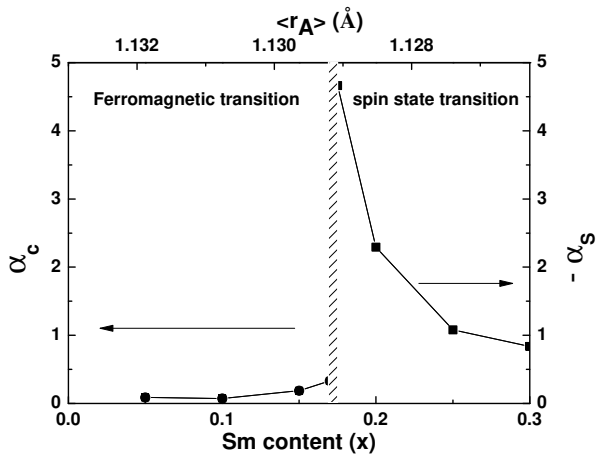}
\caption{\label{fig:epsart}  The oxygen isotope exponent
$\alpha_c$ and $\alpha_s$ as a function of Sm content x and the
average ionic radius $<r_A>$  of A site
$(Pr_{1-x}Sm_x)_{0.7}Ca_{0.3}$ in
$(Pr_{1-x}Sm_x)_{0.7}Ca_{0.3}CoO_3$ system.  }
\end{figure}
In the insulating regime (x$>$0.175), the up-shift of $T_S$
induced by oxygen isotope exchange can be explained to originate
from that substitution of $^{16}O$ by $^{18}O$ leads to a decrease
in frequency of phonon, and consequently enhances effective mass
of electron ($m^*$) through strong electron-phonon coupling. The
enhancement of $m^*$ leads to a decrease of bandwidth W, resulting
in an increase of energy difference $\Delta E$ between IS and LS
states and consequently an up-shift of $T_s$. The increase of
Sm-doping leads to a decrease in the ionic radius of A site
$(Pr_{1-x}Sm_x)_{0.7}Ca_{0.3}$ in the
$(Pr_{1-x}Sm_x)_{0.7}Ca_{0.3}CoO_3$ system, so that volume of the
$CoO_6$ octahedra decreases. The decrease in volume of the $CoO_6$
octahedra results in an increase of crystal field splitting energy
$\Delta_C$ and $\Delta E$, causing an increase in $T_S$.
Therefore, decrease of the isotope component $\alpha_s$ with
increasing x could be due to the increase of $T_s$.  In contrast
to the giant oxygen isotope effect in the insulating phase, a
negligible oxygen isotope effect on $T_c$ is observed in the
ferromagnetic metal regime. It has been reported that substitution
of $^{16}O$ by $^{18}O$ leads to a giant down-shift of Curie
temperature in $La_{1-x}Ca_xMnO_{3+y}$.\cite{zhao} Such giant
oxygen isotope shift was explained by the JT polaron formed due to
strong Jahn-Teller effect.\cite{zhao} Recently, elastic and
inelastic neutron scattering showed that spin activation of
$Co^{+3}$ ions in perovskite cobalt oxides induces local static JT
distortions in the insulating phase, while the static JT
distortions are absent in the ferromagnetic metal
phase.\cite{louca1} This observation explained that the giant
oxygen isotope effect is observed in the insulating phase, while
negligible in the ferromagnetic metal phase well. The lattice
distortions from the JT effect with strong electron-phonon
coupling should be responsible for the giant isotope effect in the
insulating phase.

In conclusion, oxygen isotope effect is systematically studied in
($Pr_{1-x}Sm_x)_{0.7}Ca_{0.3}CoO_3$ with a crossover at
x$\sim$01.75 from a ferromagnetic metal to an insulating phase
with SST. A giant oxygen isotope effect is observed in the
insulating phase with an isotope component $\alpha_S=-4.7$ for
 the sample with x=0.175 which is significantly larger than that found for any magnetic or
electronic phase transition in other oxides, while a negligible
oxygen isotope effect occurs in the ferromagnetic metal phase.
Beyond the expectation, substitution of $^{16}O$ by $^{18}O$ leads
to a phase transition from a ferromagnetic metal to an insulator
in the sample with x=0.172 which lies in the ferromagnetic metal
phase close to the crossover boundary. These contrasting behaviors
are in support of the observation that static JT distortions
occurs in the insulating phase, while are absent in the
ferromagnetic metal phase.\cite{louca1} It provides a direct
demonstration of the important of lattice vibrations in these
materials.

{\bf Acknowledgement:} This work is supported by the Nature
Science Foundation of China, and by the Ministry of Science and
Technology of China (973 project No:2006CB601001), and by the
Knowledge Innovation Project of Chinese Academy of Sciences.

\end{document}